\documentclass[11pt,twoside]{article}

\usepackage{asp2004}
\usepackage{epsf}
\usepackage{psfig}
\usepackage{lscape}

\begin{document}
\title{Properties and origin of the old, metal rich, star cluster, NGC 6791}   
\author{Giovanni Carraro}   
\affil{ESO, Alonso de Cordova 3109, 19001, Santiago de Chile, Chile}    

\begin{abstract} 
In this contribution I summarize the unique properties of the old, metal rich, star cluster NGC 6791, with particular emphasis
on its population of extreme blue horizontal branch stars.  I then conclude providing my  personal view on the origin
of this fascinating star cluster. 
\end{abstract}

\section{NGC 6791 properties}   
NGC 6791 is a  star cluster traditionally considered an open cluster following the pioneering study of Kinman (1965).
With its unique combination of  age and metallicity ($\tau$ $\approx$ 8 Gyrs and [Fe/H] $\approx$ +0.4, Carraro et al. 2006) NGC 6791 has been challenging
us for several decades. 
It is located within the solar ring, at a Galacto-centric distance of $\approx$ 7 kpc, and 1 kpc above the Galactic plane.
Being located inside the solar ring and having such a high metallicity, this cluster obviously plays a crucial role in shaping the old 
disk  radial abundance gradient (Carraro et al. 2006, Magrini et al. 2009).
Its metallicity have been measured several times. Independent estimates (Carraro et al. 2006; Origlia et al. 2006; Gratton et al. 2006; Geisler et al. 2013) 
nicely converge  to a super solar metallicity, around $[Fe/H] $+0.4. The $[\alpha/Fe]$ ratio, however, is slightly super-solar.\\

\noindent
Over the years this cluster posed several challenges. \\

\noindent
\begin{description}

\item $\bullet$ Bedin et al (2008a) obtained very deep HST  photometry of NGC 6791 down to white dwarf  (WD) cooling sequence, and firstly pointed out that the age derived fitting the WD LF was
half the age estimate from the turn off  (TO) of  single stars. This puzzle was then solved by the very same authors  (Bedin et al. 2008b) by realizing that up to 30\% of the stars in NGC 6791 are most probably binaries.

\item $\bullet$ The red giant branch (RGB) of NGC 6791 (see Fig.~1) is broad in colors (Janes 1984), yet no study detected abundance variation in Fe, 
which would be the most obvious explanation. A spread in age has been on the other hand
claimed for by Twarog et al. (2011) based on the width of the main sequence (MS)  close to the cluster TO, which was interpreted as an evidence of extended star formation.
Variable reddening, while significantly reducing this width, is not able to account for it completely (Brogaard et al. 2012).

\item  $\bullet$ The most striking result is surely the one recently presented by  Geisler et al. (2012),  who performed a detailed abundance analysis of a sample of RGB stars in the cluster.
They found that NGC 6791 does host two distinct stellar populations in Na, and, most intriguingly, the same Na-O anti-correlation as
routinely found in virtually all globulars (see Fig.~2).  

\end{description}

\noindent
This way NGC 6791 becomes the first open cluster to show an intrinsic dispersion in any element and the first presumed 
open cluster  discovered with multiple populations and/or extended star formation.
It is also the first cluster of any kind to show Na-poor stars with a homogeneous Na content, along with an Na-rich group showing an intrinsic Na scatter. 
The Na-poor group falls near the field star O/Na content, while the Na-rich population follows the NaÐO anti-correlation typical of globulars
NGC 6791 defies the traditional definition of either an open or a globular cluster.

\begin{figure}
 \plotone{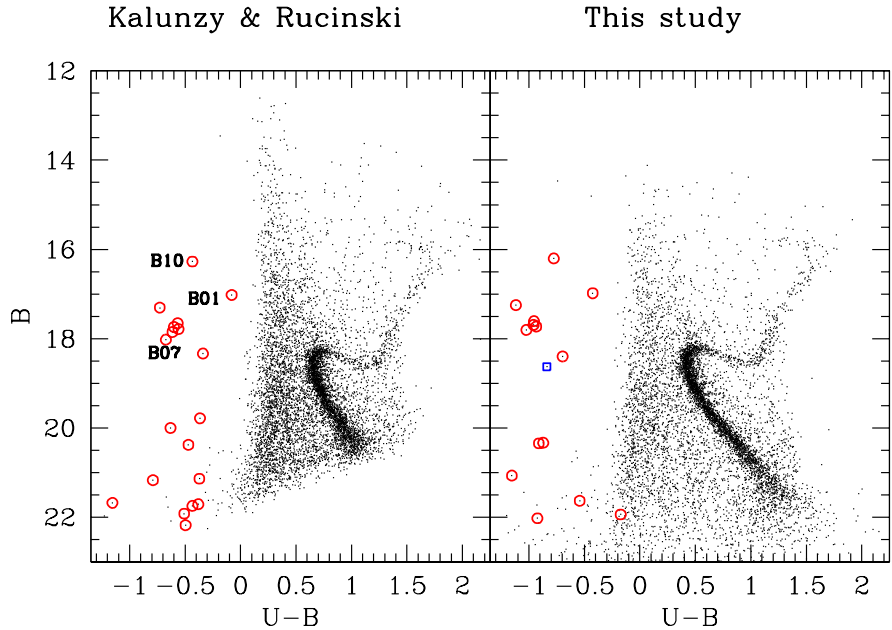}
 \caption{UB Color Magnitude Diagram of NGC 6791 from Kaluzny \& Rucinski (1995, left panel) and Carraro et al. 2013 (right panel). Red circles indicate Extreme Blue Horizontal Branch stars
 from Kaluzny and Udalski (1992), while the blue square is a new discovered candidate.}
 \end{figure}

 \begin{figure}[!h]
 \plottwo{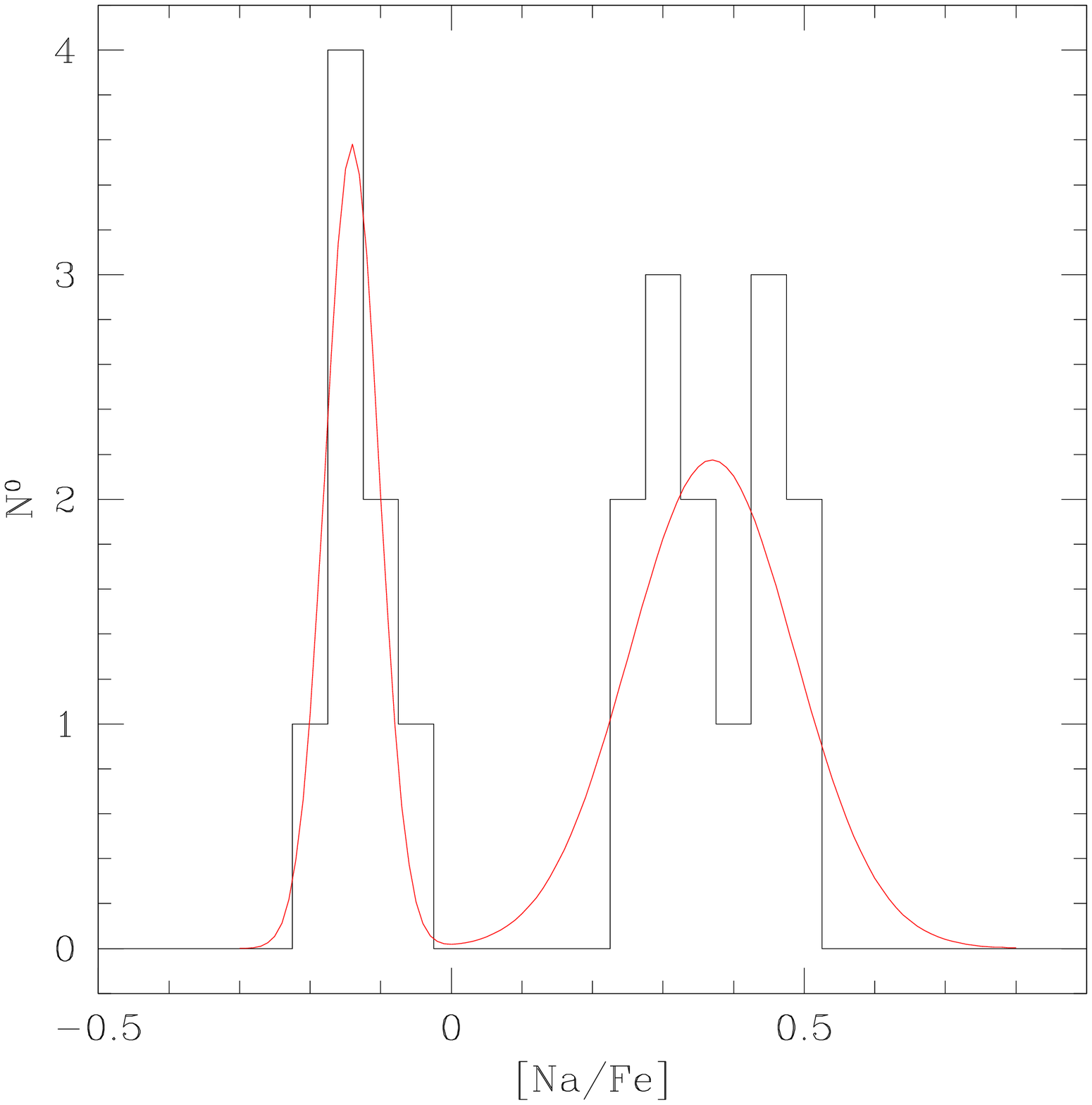}{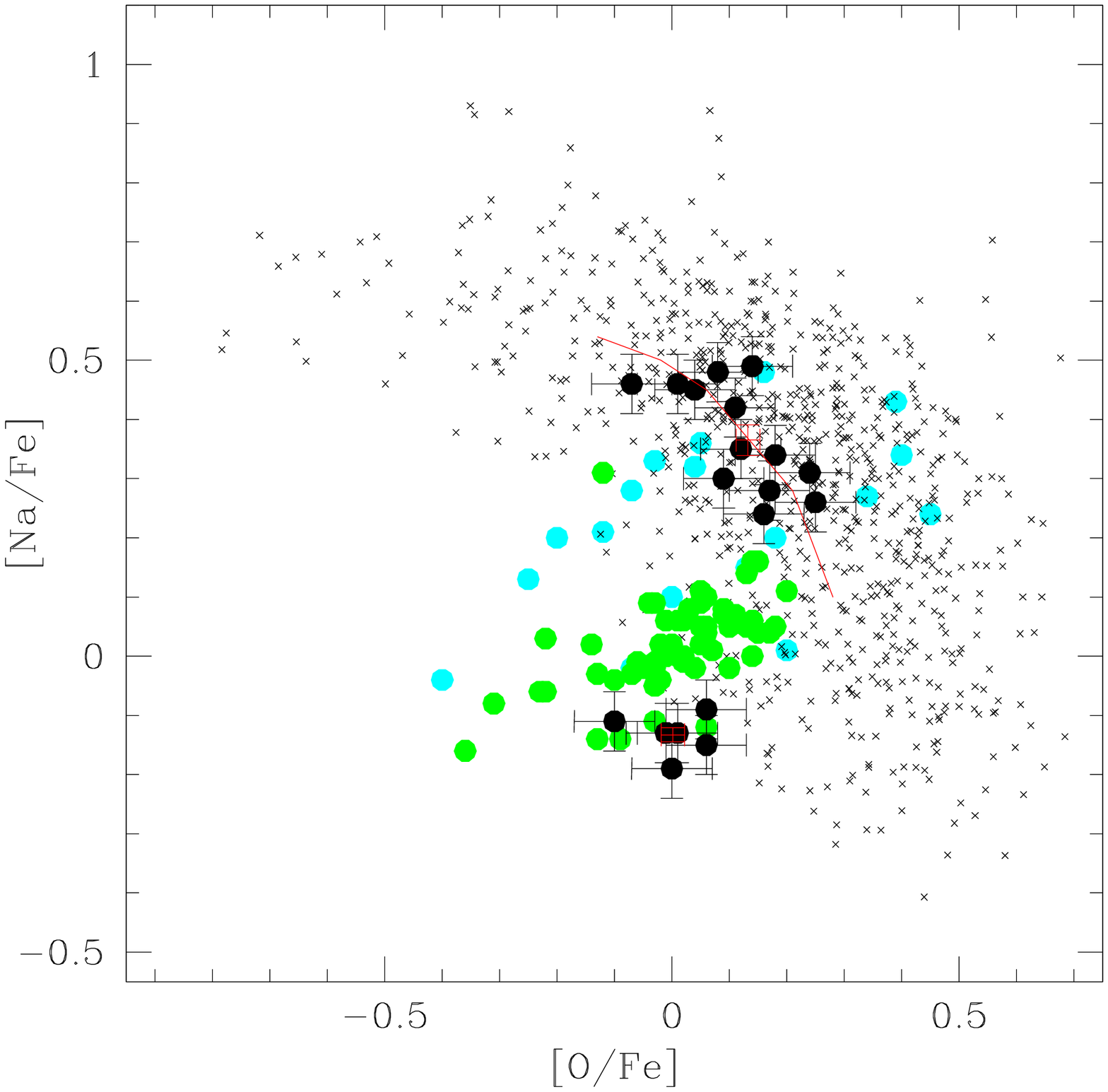}
 \caption{Left: histogram of the [Na/Fe] abundance ratio distribution (lines) with a two-Gaussian fit (curves). Right: [Na/Fe] vs. [O/Fe] for stars in NGC 6791 (filled
circles with error bars), GC stars (crosses), metal-rich ([Fe/H] $>$-0.2) field stars (green filled circles), and the means for OCs from De Silva et al. (2009) (blue filled
circles). The mean GC anti-correlation is shown by the red curve.}
 \end{figure}

\section{The population of extreme blue Horizontal branch stars in NGC 6791}   
It is known since more than 20 years that NGC 6791 harbors a  population of extremely blue stars in the magnitude range $16-18$ in B (see Fig.~1).
Kaluzny \& Udalski(1992) firstly provided a compilation of photometric candidates, which have then been followed up spectroscopically by Liebert et al. (1994).
This beautiful work already contains almost all the relevant information on this population. \\

\noindent
NGC 6791 appears to host a double horizontal branch (HB) phase. 
Most He-burning stars position in the red clump, where they have settled in after removing core degeneracy via the He-flash. This is the expected position from theory for 
such a high metallicity environment. 
However, about 20\% of He-burning stars are located in a second clump (see Fig~1) blue-ward the TO. These stars are burning He in  degenerate cores
and are surrounded by a thin envelope of masses as low as  of 0.01 $M_{\odot}$.
The narrow range in colors of these stars suggest that they  could not loose their envelope via classical mass loss along the RGB.  Empirically, this is confirmed
by the absence of dust features in RGB stars envelope (van Loon et al. 2008) and by astro-seismology in the NGC 6791 Kepler field (Miglio et al. 2012).\\

\noindent
From a theoretical point of view, mass loss is also ruled out, since standard Reimers ( $\eta \sim 0.3$) values are enough to explain the star distribution (counts)  
in the color magnitude diagram (Carraro et al. 1996).
This has been confirmed  by the recent study of Buzzoni et al.(2012).\\
\noindent
As earlier suggested by Liebert et al. (1994) and Carraro et al. (1996), a possibility for their origin is that 
they are or were in close binary systems, and they lost most of their envelope via mass transfer.
This scenario is supported by the high binary fraction detected in NGC 6791 (Bedin et al. 2008b)  and by the fact that a few BHB are still in binary systems (Liebert et al. 1994).
While very appealing, this possibility  needs further investigation, since, again, it is very hard  also in this case to reconcile the extremely narrow color range of BHB stars, which would
have been the result of an extremely fine-tuned mass loss process.\\

\noindent
Recently, one more candidate has been detected photometrically  by Carraro et al. (2013., see Fig.~1).
This study  (together with Buzzoni et al. 2012) lends support to the early suggestions by Liebert et al. (1994) 
that these stars are the responsible for the UV upturn observed  in nearby elliptical galaxies.
In particular, Buzzoni et al. (2012) performed a details stellar population synthesis study of NGC 6791 and derived its spectral energy distribution. In the blue
regimes, short-wards 2200 \AA, the flux is demonstrated to be entirely produced by BHB stars.

\section{The origin of NGC 6791}
Elliptical galaxies in the local volume with the same metallicity as NGC 6791 have masses around $10^{11}-10^{12}$ $M_{\odot}$(Gallazzi et al. 2006). 
It is therefore difficult to conceive a scenario where NGC 6791 had lost almost all its mass while engulfing in the Milky Way to get nowadays a mass around $5\times 10^3$ $M_{\odot}$(Kinman 1965).\\

\noindent
A different scenario has to be looked for.\\
\noindent
In Jilkova et al. (2012) we investigated the possibility that NGC 6791 formed close to the bulge, where, supposedly, star formation has been strong and chemical enrichment fast.
Having formed there, a plausible migration mechanism has to be searched for to explain the actual cluster location.\\

\begin{figure}[!h]
 \plotone{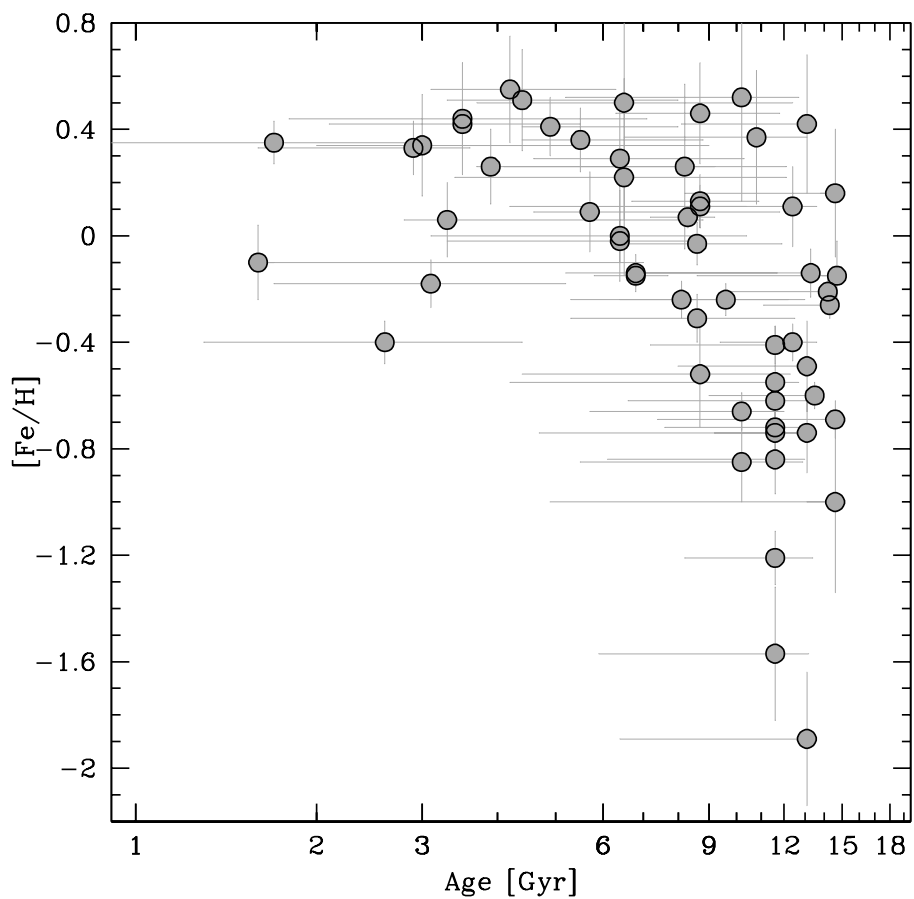}
 \caption{Age metallicity relation for micro-lensed dwarf from Bensby et al. (2013). Notice how NGC 6791 nicely fit into this relationship with an age of 8 Gyr and a metallicity [Fe/H]=+0.4}
 \end{figure}

\noindent
I would like to argue here that a formation close to the bulge is really an appealing and reasonable scenario.  In Fig.~3 we show the age metallicity  relationship from a sample
of micro-lensed dwarfs studied by Bensby et al (2013).  The age (8 Gyr) and metallicity ([F/H]=+0.4) of NGC 6791 is totally compatible 
with the trend defined by micro-lensed  stars in the bulge. \\

\noindent
Jilkova et al. (2013) investigated the orbital parameters of test particles moving close to the Galactic bulge under the influence of a Galactic potential which includes perturbations
from the Galactic bar and spiral arms. The underlying motivation was to look for orbits similar to the actual one for NGC 6791. The orbital parameter for NGC 6791 were  previously determined integrating back in time
its orbit for a short time of about 1 Gyr. In Fig.~4 we show a successful orbit, namely an orbit which brings a test particle from the inner disk to the actual position of NGC 6791. The orbital parameters are
amazingly similar to the ones of NGC 6791.\\

\noindent
A statitical analysis performed on over 1,000 orbits shows, however, 
that this particular orbit has, under the adopted potential, less than 10\% probability to occur. This has been previously interpreted as an indication
that the formation of NGC 6791 close to the bulge and its later displacement outwards is highly improbable.\\

\noindent
I argue here, instead, that this is a plausible scenario, since {\it we do not know any other star cluster with the same properties as NGC 6791}.\\

\noindent
While surely further studies are, as usual, needed, I am keen to consider NGC 6791 as a prototype of the Galactic bulge population, and not a genuine disk star cluster. It is so unique that considering it as an open cluster is not anymore possible.  Deriving global properties of the Galactic disk including NGC 6791 is therefore not recommended.\\

 \begin{figure}
 \plotone{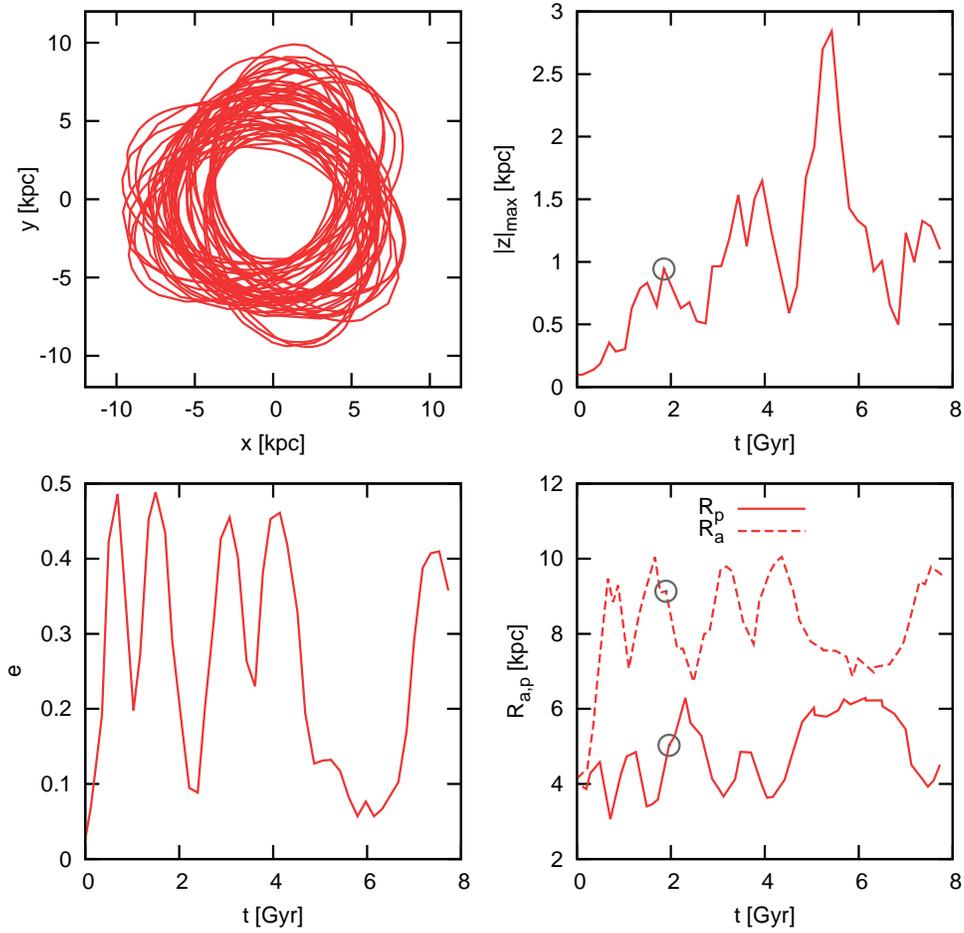}
 \caption{Example of a galactic orbit that reproduces the actual orbital parameters for NGC 6791. This orbit has been calculated under the influence of a time independent galactic potential
 which, however, includes perturbations from the bar and spiral arms..}
 \end{figure}

\noindent 
{\it I would like to acknowledge the many colleagues I worked with over the years in this project,
in particular Alberto Buzzoni, Lucie Jilkova, Sandro Villanova, and Doug Geisler. }


\begin{thebibliography}{}
\bibitem[\protect\citeauthoryear{Bedin et al.}{2008}]{bedin08a} Bedin L.R., King, I.R., Anderson, J., Piotto, G., et al., 2008a, ApJ, 678, 1279
\bibitem[\protect\citeauthoryear{Bedin et al.}{2008}]{bedin08b} Bedin L.R., Salaris M., Piotto, G., Cassisi S., et al., 2008b, ApJ, 678, 1279
\bibitem[\protect\citeauthoryear{Bensby et al.}{2013}]{bensby13} Bensby T., Yee J.C., Feltzing S., Johson J.A., et al., 2013, A\&A, 459, 147
\bibitem[\protect\citeauthoryear{Brogaard et al.}{2012}]{brogaard12} Brogaard J., Vandenberg D.A., Bruntt H., Grundahl F.,  et al., 2012, A\&A, 543, 106
\bibitem[\protect\citeauthoryear{Buzzoni et al.}{2012}]{buzzoni12} Buzzoni A.,Bertone E., Carraro G., \& Buson, L.M., 2012, ApJ, 749, 35
\bibitem[\protect\citeauthoryear{Carraro et al.}{2006}]{carraro06} Carraro G., Villanova S., Demarque P., McSwain M.V., Piotto G., \& Bedin L.R. 2006, ApJ, 643, 1151
\bibitem[\protect\citeauthoryear{Carraro et al.}{2013}]{carraro13} Carraro G., Buzzoni A., Bertone E.,  \& Buson L.M., 2013, AJ, in press
\bibitem[\protect\citeauthoryear{Gallazzi et al.}{2005}]{gallazzi06} Gallazzi A., Charlot S., Brinchmann J., White S.D.M., \& Tremonti C.A., 2005, MNRAS, 362, 41
\bibitem[\protect\citeauthoryear{Geisler et al.} {2012}]{geisler12} Geisler D., Villanova S., Carraro G., Pilachowski C., Cummings J., Johnosn C.I.,  \& Bresolin F., 2012, ApJ, 756, L40
\bibitem[\protect\citeauthoryear{Gratton et al.}{2006}]{gratton06}Gratton R., Bragaglia A., Carretta E., \& Tosi M., 2006, ApJ 642, 462
\bibitem[\protect\citeauthoryear{Janes}{1984}]{janes84}Janes, K.A., 1984, PASP, 96, 977
\bibitem[\protect\citeauthoryear{Jilkova et al.}{2012}]{jilkova12}Jilkova L., Carraro, G., Jungwiert B., \& Minchev I., 2012, A\&A , 541, 64
\bibitem[\protect\citeauthoryear{Kaluzny \& Rucinski}{1995}]{kaluzny95} Kaluzny, J., \& Rucinski S.M. 1995, A\&AS, 114, 1
\bibitem[\protect\citeauthoryear{Kaluzny \& Udalski}{1992}]{kaluzny92}  Kaluzny, J., \& Udalski, A. 1992, AcA, 42, 29
\bibitem[\protect\citeauthoryear{Kinman}{1965}]{kinman65} Kinman, T.D. 1965, ApJ, 142, 655
\bibitem[\protect\citeauthoryear{Liebert et al.}{1994}]{liebert94} Liebert, J., Saffer, R.A., \& Green, E.M. 1994, ApJ, 107, 1408
\bibitem[\protect\citeauthoryear{Magrini et al.}{2010}]{magrini09} Magrini L., Sestito P., Randich S., Galli, D., 2009, A\&A, 494, 95
\bibitem[\protect\citeauthoryear{Miglio et al.}{2012}]{miglio12} Miglio A., Brogaard, K., Stello, D., Chaplin, W.J., D'Antona, F., et al., 2012, MNRAS, 419, 2077  
\bibitem[\protect\citeauthoryear{Origlia et al.}{2006}]{origlia06} Origlia L., Valenti E., Rich R.M., \& Ferraro F.R. 2006, ApJ, 646, 499 
\bibitem[\protect\citeauthoryear{Twarog et al.}{2011}]{twarog11} Twarog B.~A., Carraro G., \& Anthony-Twarog B.~J., 2011, ApJ, 727, L7 
\bibitem[\protect\citeauthoryear{van Loon et al.}{2008}]{vanloon08} van Loom J. Th., Boyer, M.L., McDonald, I., 2008, ApJ, 680, L49
\end{thebibliography}
\end{document}